Raman study of the Verwey transition in Magnetite at high-pressure and low-temperature; effect of Al doping.

L. Gasparov[1], Z. Shirshikova[1], T. M. Pekarek[1], J. Blackburn[1], V. Struzhkin[2], A. Gavriliuk[2,3,4], R. Rueckamp[5], H. Berger[6]

[1] University of North Florida, Jacksonville, Fl

[2] Geophysical Laboratory, Carnegie Institution of Washington, Washington, DC

[3] Institute for High Pressure Physics, Russian Academy of Sciences, Troitsk, Russia

[4] Institute of Crystallography, Russian Academy of Sciences, Moscow, Russia

[5] University of Cologne, Institute of Physics 2, Cologne, Germany

[6] Ecole Polytechnique Federale de Lausanne, Switzerland

We report high-pressure low-temperature Raman studies of the Verwey transition in pure and Al –doped magnetite ($Fe_3O_4$) that employ neon as a pressure transfer medium. Below the transition Raman spectra of magnetite display a number of additional phonon modes that serve as transition markers. These markers allow one to investigate the effect of hydrostatic pressure on the transition temperature. Al-doped magnetite $Fe_{2.8}Al_{0.2}O_4$ ($T_V$=116.5K) displays a nearly linear decrease of the transition temperature with an increase of pressure yielding $dP/dT_V$=-0.096±0.013 GPa/K. Pure magnetite displays a significantly steeper slope of the PT equilibrium line with $dP/dT_V$ = -0.18±0.03 GPa/K. Contrary to earlier high pressure resistivity reports we do not observe quantum critical point behavior in pure magnetite at 8 GPa. We compare our spectroscopic data with that obtained from the ambient pressure specific heat measurements and find a good agreement in pure magnetite. Our data indicate that Al doping leads to a smaller entropy change and larger volume expansion at the transition. Our results are consistent with the mean field model of the transition that assumes charge ordering.

## I. INTRODUCTION

### A. Verwey transition in Magnetite

Magnetite ($Fe_3O_4$) is naturally occurring mineral. It is the first magnetic material known to mankind.[1-2] In spite of decades of research, comprehensive understanding of this compound is lacking. In particular, the Verwey transition[3] in magnetite discovered in 1939 remains unexplained. At ambient pressure the Verwey transition in pure or near-stoichiometric magnetite is of a first-order. The transition occurs at $T_v \sim 123$ K, with changes in crystal structure, latent heat, and a two-order of magnitude decrease in dc-conductivity. Oxygen, iron deficiency, and/or doping may reduce the transition temperature and cause the transition to become higher order, or suppress it completely.

Above the transition, magnetite has cubic inverse spinel structure[3-6] (*F3dm*) with two distinct iron sites. The so-called A-site has tetrahedral oxygen coordination and the B sites are octahedrally coordinated. The A-sites are occupied by $Fe^{3+}$ ions only, whereas the B-sites can host $Fe^{+2}$ and $Fe^{+3}$ ions. Below the transition, the crystal structure is likely monoclinic with orthorhombic *Pmca* pseudosymmetry.[7-9]

To this day none of the proposed mechanisms[10-16] including original Verwey-Haayman charge ordering model of the transition[10] successfully describe the whole body of experimental data. The issue of charge ordering at the Verwey transition remains far from settled. Recent X-ray, neutron, and electron diffraction experiments[9,17-19] have cast considerable doubt on both the Verwey order-disorder model, and Anderson's model.[11] However, resonance X-ray diffraction data of Nazarenko et al.[20] and Joly et al.[21] confirmed partial charge ordering at the transition.

## B. Doping

Coulomb interaction likely plays an important role in the mechanism of the Verwey transition. Any perturbation that changes the ideal stoichiometry of the magnetite will alter the $Fe^{+2}/Fe^{+3}$ ratio on the octahedral (B) sites. Substitution of the iron ions by metal dopants creates such perturbation and therefore is of interest in the understanding of the transition mechanism. A number of publications[22-26] report the studies of magnetite's doping with Zn, Ti, Al, Ga, Ni, Co, Mg and Cr. Aluminum substitution presents an interesting case since Al ions can substitute both A- and B- sites in the magnetite unit cell[22], whereas Ti and Zn ions enter B sites only. Substitution of the A-sites has a minor effect on the transition whereas B-sites play a much more prominent role in its mechanism. Since both A and B sites can accept Al it requires higher Al concentration to yield the effect similar to Zn and/or Ti doping.

## C. Effect of Pressure

Application of hydrostatic pressure is a clean way of affecting the transition, since no change of chemical composition takes place. The transition temperature decreases with pressure,[27-36] however the reported rate of such decrease differs significantly from one report to the other. The most recent high pressure measurements of pure magnetite[33,35-36] suggest the presence of a discontinuity of the transition. In particular the high pressure resistivity measurements of Mori et al.[33], resistivity and magnetization measurements of Spalek et al.[36] point to a metallization of magnetite at 0K and above 8 GPa. A similar behavior is suggested by recent high pressure low temperature x-ray powder diffraction (XRPD) study of Rozenberg et al.[35], however, with only four phase transition points detected and no data taken below 82K the suggested disappearance of the transition above ~8GPa is debatable.

A number of problems may affect the high pressure data extracted from the transport measurements. The transport measurements are inherently affected by the quality of electric contacts. The transport properties are governed by the interplay of carrier concentration, which is a function of the density of states, and mobility that depends upon a number of parameters including defect concentration. The authors of Ref. 36 employed annealing of the samples with subsequent quenching to room temperature that could have potentially lead to increase of defect concentration in the crystals.

Raman scattering has certain advantages compared to transport- or magnetization- based techniques. Raman scattering does not suffer from the problem of electric contacts. It is sensitive to changes in lattice dynamics and therefore it is especially useful for the study of the Verwey transition where the crystal symmetry change is one of the major indicators of the transition. Raman spectra of magnetite below the transition display additional phonon modes appearing due to increased unit cell.[4-8] These spectral lines can serve as transition markers, enabling one to investigate how the Verwey transition temperature changes with pressure.

II. EXPERIMENTAL

Single crystals of pure and doped $Fe_3O_4$ have been grown by standard chemical vapour phase method by using HCl as the transport gas for the crystal growth.

Corresponding endothermic reaction is: $Fe_3O_4$ (s) + 8HCl (g) → $FeCl_2$ (g) + $FeCl_3$ (g) + $4H_2O$ (g)

Aluminium doping was accomplished by adding Al oxide powder to the $Fe_3O_4$ charge. The tubes were positioned in a horizontal two zone tube furnace with a temperature gradient of approximately 950°C to 800°C for 14–21 days, with the charge in the hot end of the tube. This method yielded the samples of a typical size of 4x4x2 mm

X-ray diffraction confirmed the spinel type structure of the crystals. Electron microprobe was used for initial chemical analysis. The pure magnetite single crystal was characterized by resistivity measurements and displayed a sharp transition at 123K. High pressure measurements of this sample have been previously published.[34]

The Al-doped magnetite sample ($Fe_{3-x}Al_xO_4$) was first studied by room temperature Raman scattering in order to confirm its homogeneity. We measured Raman spectra in 20 different points on the sample. We were particularly interested in the two most prominent modes in the spectrum[37-43] at around 670 cm$^{-1}$ and 540 cm$^{-1}$ that are associated with oxygen vibrations in the Fe-O tetrahedron (A-site).[37-43] The variations of the frequencies of these modes would indicate inhomogeneity of our Al-doped sample. We found that the $A_{1g}$ mode was centered at 668.3 cm$^{-1}$ with a standard deviation of only 1.35 cm$^{-1}$. The width of the mode was found to be 39.8 cm$^{-1}$ with a 0.8 cm$^{-1}$ standard deviation. The $T_{2g}$ was centered at 540.5 cm$^{-1}$ with a standard deviation of 1.4 cm$^{-1}$. Such a small deviation of the frequency is well within instrumental error.

The Al –doped sample was split into several pieces. Two of them were used to measure the specific heat and magnetization on two different PPMS setups and a SQUID setup. The third piece was used to provide microscopic samples for the high pressure Raman measurements. To determine the Al concentration, the largest piece of the sample was sent to the Columbia Analytical Services Inc. where it was dissolved to determine the Al concentration using titrimetry. The latter method yielded 2160 ppm of Al in the ($Fe_{3-x}Al_xO_4$) sample which translates into x=0.019.

The specific heat measurement of the Al-doped sample displays a peak at 119 K with about a 3K full width at half maximum, Fig. 1. The magnetization measurements, Fig. 1, indicate the onset of the transition at 121 K with a transition width of about 9K. The transition displayed about a half a degree hysteresis. Following Miyahara[22] we determined the Verwey transition temperature to be in the center between the onset of the magnetization drop and the end of the drop yielding $T_V$=116.5 ± 4.5K. The value of the Verwey temperature and the width of the transition indicate that the transition is of the second order. The transition temperature, corresponding Al concentration, and the order of transition are in agreement with previously published data.[21-24] One should note a significant uncertainty in determining the transition temperature. In particular, had the onset of magnetization transition been used, the transition temperature would be at 121K. Whereas the peak of the specific heat would yield $T_V$=119K. The large spread of the values of $T_V$ as a function of Al content[21-24] is expected since doping can lead to both Fe ion substitution and the creation of iron vacancies with both effects affecting the transition.

The high pressure experiment employed a miniature Diamond Anvil Cell developed in the Geophysical Laboratory.[44] To maintain hydrostatic pressure the sample of about 15μm was placed into the opening of a gasket between two synthetic diamond anvils. For the Al-doped sample we used neon as a pressure transfer medium. The laser light was shone through the diamonds on the sample and Raman scattering was collected. The laser power did not exceed 6 mW in front of the cryostat's window. No significant sample overheating was detected for the Al-doped sample. The magnitude of the pressure was measured by the shift in frequency of the ruby luminescence line.

III. RESULTS AND DISCUSSION

Fig.2 displays the results of a typical high pressure run. At 6.4 GPa pressure, as temperature increased above 45K, one could clearly see the disappearance of the spectral markers associated with the low-temperature phase of magnetite, indicating that the Verwey transition took place. From these data one can suggest that $T_V$ decreased from 120K at atm. pressure to about 45K at 6.4GPa. The summary of such runs is displayed in Fig.3. We could detect the low temperature phase of magnetite all the way up to 8 GPa. The change of the transition temperature is linear with the rate of about 10K per GPa or $dP/dT_V$=-0.096 ±0.013 GPa/K. At 8 GPa we could not detect a transition to the low temperature phase down to 5 K. We would like to note that 10K per GPa decrease of the transition temperature in the Al doped magnetite is significantly larger than any published suppression rate in a pure magnetite[27-36] indicating that Al doping strongly affects pressure dependence.

In Fig.3 we contrast the data obtained from the Al-doped sample with our previously published pure magnetite data.[34] When we performed these measurements in 2005 we initially used neon as a pressure transfer medium. When attempting lower pressure experiment we discovered a substantial overheating of the sample. To address this problem, we had to defocus the laser which allowed us to measure 20 GPa, 30K phase transition point. To avoid overheating and laser defocusing at higher temperature and lower pressure we put the sample in direct contact with the diamond anvil and used NaCl as the pressure transfer medium resulting in about 2 GPa pressure gradient over the different areas of the sample. However the highest pressure data point (20GPa) did not suffer from potential pressure gradient since it was measured with neon as pressure transfer medium. This observation is in stark contrast with the reports indicating metallization of magnetite above 8GPa. The experiment[34] yielded the transition temperature decrease of 5.6 K/GPa or $dP/dT_V$ =-0.18±0.03 GPa/K. In Fig.3 we compare the results of our Raman study with the high pressure XRPD data of Rosenberg et al.[35] This data do not extend below 82K. The data is in a very good agreement with our Raman results. We would like to note that quantum critical point behavior in the pure magnetite above 8GPa suggested by the Refs. 33, 35-36 is not supported by our data.

Pressure- Temperature (P-T) phase separation line provides valuable information about thermodynamic parameters of a phase transition. In particular, the Clausius–Clapeyron formula relates the slope of the P-T phase separation line with the change of molar entropy and molar volume at the transition. i.e. $dP/dT_V = \Delta s/\Delta v$, where $\Delta s$ is the change of the molar entropy of the transition and $\Delta v$ is the change of molar volume at the transition. The comparison is straightforward in pure magnetite. In particular, experimentally observed volume expansion of the magnetite at the transition[8] is $\Delta V/V = 6*10^{-4}$. Molar volume of magnetite can be calculated from its molar mass of 0.23154 kg/mol and its average density of 5150 kg/m$^3$ leading to a molar volume of $4.50*10^{-5}$ m$^3$/mol. This in turn gives the change of molar volume at the transition equal to $2.7*10^{-8}$ m$^3$/mol. This change of volume is actually negative since magnetite expands when going from the high temperature phase to the low temperature phase.[7-9]

Shepherd et al.[46] reported the change of the molar entropy at the Verwey transition obtained from the specific heat measurements. The averaged value of $\Delta s$ over nine measurements of three magnetite samples with $T_V$ between 120.5 and 121.1K is $\Delta s = 5.91 \pm 0.06$ J/(mol*K). After entering these data in the Clausius-Clapeyron formula we obtain $dP/dT_V = -0.22$ GPa/K, a value which is in excellent agreement with our experimental value of $-0.18 \pm 0.03$ GPa/K.

Specific heat measurements of our Al-doped sample ($T_V=116.5$K) yielded the average value of the molar entropy change of $4.66 \pm 0.10$ J/(mol*K). We are not aware of the data for the molar volume change at the Verwey transition in the Al doped magnetite. If we instead use the molar volume change in a pure magnetite we will arrive to the $dP/dT = -0.17$ GPa/K. This value is nearly double the experimentally observed value -0.096±0.013 GPa/K. Most likely reason for such discrepancy is the fact that the molar volume change for the Al-doped sample is likely different than that of a pure magnetite. Remarkably, the relative volume change of $\Delta V/V=1.5*10^{-3}$ reported by Rozenberg et al.[35] for their magnetite powder would be in agreement with our data for Al-doped single crystal of magnetite. Furthermore, we do observe disappearance of the transition above 8GPa in the Al-doped sample but not in the pure magnetite. The third law of thermodynamics predicts that the entropy of a system approaches zero at zero temperature. The latter means that the PT phase separation line should flatten since the Clausius-Clapeyron formula will have a vanishing nominator and will result in zero $dP/dT_V$. This may be the reason why we do not see a transition at 8GPa. It is plausible the transition does happen, however, at the temperatures that are below the limit of our technique. A similar argument would imply that the same flattening of the PT line should occur in pure magnetite. Measurements at extremely low temperatures and high pressures could clarify this point.

One conclusion of our study is that doping strongly affects the slope of the PT phase equilibrium line, whereas the change of the Verwey transition temperature with doping remains to be just a few degrees. It is worth noting the large spread of the reported PT slopes in pure magnetite.[27-36] The reason for such spread is quite possibly the fact that the reported measurements are actually done on the magnetite samples that have significantly different iron or oxygen deficiency, or possibly doping.

The change of the PT equilibrium line slope can be understood in the following way. One expects a decrease of the molar entropy change of the transition with doping because of potential iron ion vacancies and Al substitution takes place, thus reducing the number of states participating in the transition. The trend in the change of the molar volume is less clear. One may assume that the molar volume change at the transition will likely depend on the dopant's ionic radii, thus creating a significant spread in the PT slope data that is indeed reported in the literature.[27-36]

A modified mean field model of Verwey transition[45] takes into account the volume effects. The pressure effect on the Verwey transition temperature has been described[45] as

$\Delta T_V = -\dfrac{6v_e(p-p_0)}{7Nk}$, where $v_e$ is the relative volume expansion at the transition ($v_e = \Delta V/V$), $p-p_0$ is the pressure increase relative to atmospheric pressure $p_0$, $N$ is the concentration of the unpaired 3d electrons that are responsible for ordering at the octahedral sites, and $k$ is the Boltzmann's constant. After rearranging the terms one obtains $\dfrac{dP}{dT_V} = -\dfrac{7Nk}{6v_e}$.

To account for the experimentally observed smaller compared to pure magnetite $dP/dT_V$ slope in the Al-doped sample, one would need to imply increased volume expansion at the Verwey transition and/or a decrease in the concentration of unpaired 3d electrons (N) responsible for ordering. Al-doping results in $Al^{+3}$ ion substituting $Fe^{+2}$ or $Fe^{+3}$ ions[30] and possibly iron vacancies, resulting in a decrease of N.

The trend predicted by Brabers et al.[45] is consistent with the observed decreased molar entropy change in the Al-doped sample and it is consistent with the Clausius-Clapeyron analysis we presented above. The theory[45] assumes charge ordering in magnetite and our data is in agreement with it.

IV. CONCLUSIONS

In this paper we present the high pressure measurements of as grown single crystals of magnetite using Raman spectroscopy. We argue that reported[33,35-36] discontinuity of the transition at 8GPa is not intrinsic to pure stoichiometric magnetite but rather a characteristics of a departure from the ideal stoichiometry.

In pure magnetite we find an excellent agreement between $dP/dT_V$ obtained from the Raman based high pressure experiment and that predicted by the Clausius-Clapeyron relation and the ambient pressure specific heat and volume expansion data.

Based on our data we suggest that the volume expansion at the Verwey transition should increase significantly in the Al-doped magnetite compared to a pure magnetite. Our data is in agreement with the charge ordering based mean field theory[45] of the Verwey transition.

ACKNOWLEDGEMENTS:

LG acknowledges the support from the National Science Foundation (NSF) Grants DMR-0805073, DMR-0958349, and the Office of Naval Research award N00014-06-1-0133. TMP acknowledges the support of the UNF Terry Presidential Professorship, and the NSF Grant DMR-07-06593. HB acknowledges the support of the Swiss NCCR research pool MaNEP of the Swiss NSF. We would like to thank Dr. Bussy (University of Lausanne) for the microprobe analysis. VS acknowledges the support from DOE/BES under grant No. DE-FG02-02ER45955, which was instrumental for the high pressure part of this work. AG acknowledges the support of the Russian Foundation for Basic Research grants 09-02-01527-a, 11-02-00291-a, and 11-02-00636-a, and the Russian Ministry of Science grant 16.518.11.7021

FIGURES

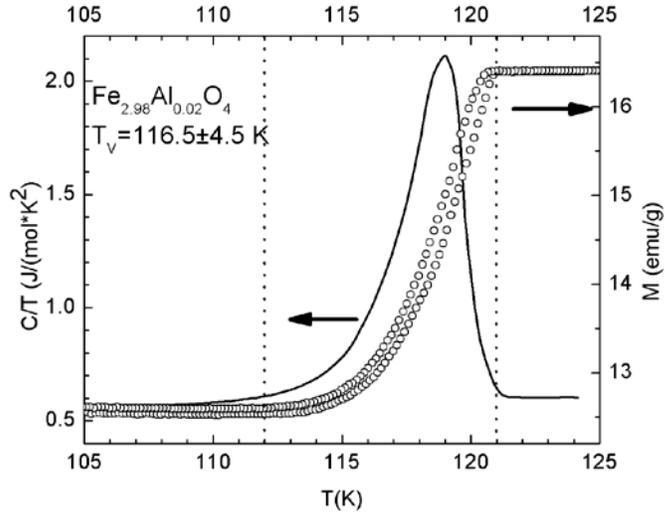

FIG. 1. Specific heat (left axis) and magnetization (right axis) data measured in the $Fe_{2.98}Al_{0.02}O_4$. Solid line indicates specific heat divided by the temperature as a function of temperature. The integral of the specific heat peak yields change of the molar entropy at the transition. Open circles indicate magnetization data. A half degree hysteresis is clearly visible in the data. Straight dashed lines indicate the width of the transition. We determined the transition temperature to be in the center between the onset of the magnetization drop and the end of the drop.

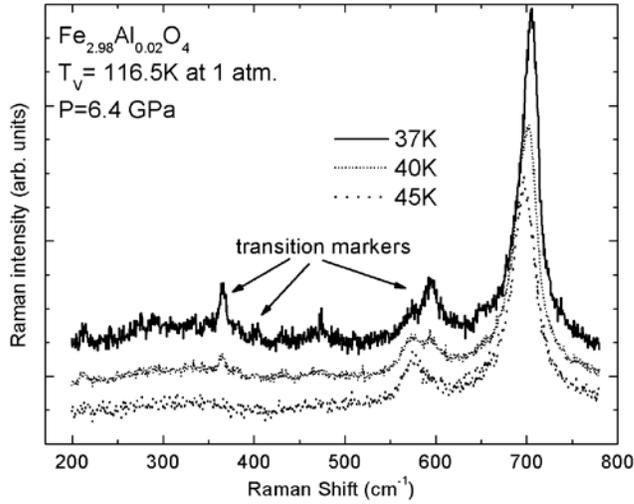

FIG. 2. Low temperature Raman spectra of $Fe_{2.98}Al_{0.02}O_4$ at 6.4 GPa. Dashed line corresponds to 45K-spectrum, dotted line corresponds to 40K-spectrum, and solid line corresponds to 37K-spectrum. Arrows indicate transition markers. The 45K-spectrum corresponds to the low temperature phase. The 37K spectrum corresponds to the high temperature phase. The 40K-spectrum corresponds to a mixed state.

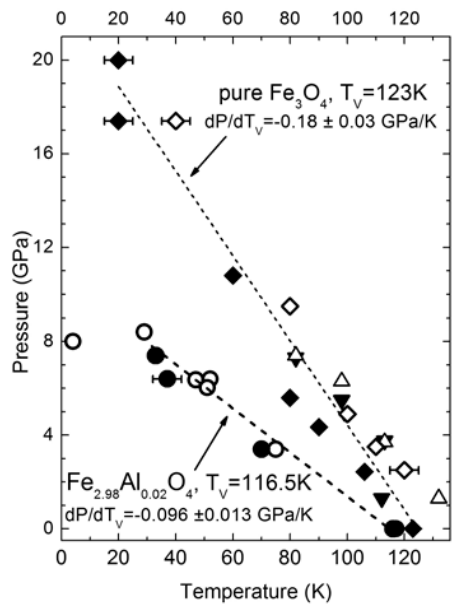

FIG. 3. Verwey transition temperature as a function of pressure. Open symbols indicate low temperature phase; filled symbols indicate the high temperature phase. Circles correspond to Al-doped sample, diamonds indicate our previously published data[34] in pure magnetite, and triangles represent the XRPD data of Rosenberg et al.[35] Error bars indicate typical uncertainty of the technique. Note that the pressure error bars are smaller than the symbols. Dashed lines indicate suggested phase separation PT equilibrium lines.